\documentclass[preprint]{elsarticle}
\usepackage{lineno,hyperref,pdflscape}
\modulolinenumbers[5]

\usepackage{multicol}
\usepackage{color}
\usepackage{xspace}
\usepackage{enumerate}
\usepackage{amsmath} 
\usepackage{ulem} 
\usepackage{placeins}
\usepackage{threeparttable}

\usepackage{tabularx}
\usepackage{mathtools} 
\usepackage{amsfonts}
\newcolumntype{M}{>{\centering\arraybackslash}m{1.85cm}}
\usepackage[export]{adjustbox}
\usepackage{float}
\usepackage{braket}
\usepackage{lipsum}
\usepackage{longtable}
\usepackage{lscape}
\graphicspath{{figure/}}
\newcommand\T{\rule{0pt}{3ex}}       
\newcommand\B{\rule[-1.5ex]{0pt}{0pt}} 

\usepackage{booktabs}
\usepackage{blindtext}
\makeatletter
\newcommand{\colorcaption}[2][]{%
	\begingroup%
	\renewcommand{\@caption@fignum@sep}{ (Color online). }%
	\caption[#1]{#2}%
	\endgroup%
}

\makeatother

\usepackage{amssymb}
\usepackage{graphicx}
\usepackage{rotating}
\usepackage{tikz}
\begin{document}
	
\begin{frontmatter}
\title{Study of structure and radii for $^{20-31}$Na isotopes using microscopic interactions}
\author{Subhrajit Sahoo$^{1}$}
\author{Praveen C. Srivastava$^{1}$,\footnote{Corresponding author: praveen.srivastava@ph.iitr.ac.in}}
\author{Toshio Suzuki$^{2}$}
\address{$^{1}$Department of Physics, Indian Institute of Technology Roorkee, Roorkee 247667, India}
\address{$^{2}$Department of Physics, College of Humanities and Sciences, Nihon University, Sakurajosui 3, Setagaya-ku, Tokyo 156-8550, Japan}

\date{\hfill \today}
\begin{abstract}

In this work, Na isotopes with mass varying from $A=$ 20 to 31 have been studied using DJ16A, JISP16 and N3LO microscopic effective interactions in the $sd$-shell. These effective interactions are derived for $sd$-shell   using no-core shell model wavefunctions and a unitary transformation method. We have also performed calculation with IMSRG effective interactions targeted for a particular nucleus. The studies include a detailed analysis of ground state binding energy, low-lying spectra, and electromagnetic properties such as reduced electric quadrupole transition strengths, quadrupole and magnetic dipole moments of the Na chain. The results obtained from these microscopic effective interactions were compared with the experimental data as well as with the results of phenomenological interaction USDB. The charge radii of Na isotopes are evaluated using shell model harmonic oscillator wave functions. In addition to charge radii, matter radii and neutron skin thickness in the Na chain are discussed with a focus on neutron-deficient Na isotopes.

\end{abstract}
\end{frontmatter}
	
	


\section{Introduction}
In the last few years, the {\it ab initio} approaches have been very successful in describing nuclear structure properties. These methods solve the nuclear many-body problem starting from the realistic two- or three-nucleon forces. Some of the {\it ab initio} approaches widely used to study nuclear structure properties are many-body perturbation theory (MBPT) \cite{mbpt_1,mbpt_2}, coupled-cluster theory \cite{cc_1,cc_2}, no-core shell model (NCSM) \cite{ab_initio_no_core_review1,ab_initio_no_core_review2},  and in-medium similarity renormalization group (IMSRG) \cite{imsrg_start,imsrg_A_dependent,imsrg_Z_dependent}. Recently, the NCSM has achieved tremendous success in nuclear structure studies, such as reproducing binding energies, low-lying spectra, transition strengths, beta decay properties etc. However, because of limited computational resources, only light or $p$-shell nuclei can be studied with these approaches. The IMSRG {\it ab initio} approach has also earned popularity in recent years for covering a wide range of the nuclear chart. It also includes the contribution of \textit{NNN} forces and can target the ground and a few excited states of light and medium mass nuclei.

Alternatively, the nuclear structure properties of the medium and higher mass nuclei can be studied in the shell model approach using different effective interactions derived through {\it ab initio} methods. In this approach, many nuclear observables, such as excitation spectra, transitions, moments, radii, deformation properties, etc., can be accessed.  Effective interactions are derived for $sd$-shell from microscopic \textit{NN} interactions such as N3LO potential inspired from chiral effective field theory \cite{N3LO}, j-matrix inverse scattering potential (JISP16) \cite{JISP16} using the NCSM approach \cite{eff_int_method(2015)}. The low-lying energy spectra of $^{18}$F calculated from these effective interactions are identical to the spectra obtained from NCSM for $^{18}$F. These effective interactions also reproduce the experimental energy spectra in closed-(sub) shell nuclei well. Inspired by the previous reference, a new effective interaction is developed for $sd$- shell from Daejon 16 potential \cite{DJ16} using the same formalism in Ref. \cite{sdshell_int}. The DJ16 interaction is further improved by making monopole modifications (DJ16A) in the same reference. The DJ16A interaction well describes O isotopes, odd-$A$ F isotopes, and some isotopes of Si and S.  The IMSRG approach is also used to derive a mass-dependent valence space (VS) effective interaction for $sd$ - shell starting from chiral next-to-next-to-next-to-leading order (N3LO) potential \cite{imsrg_A_dependent}. The energy spectra obtained for F and Ne isotopes from this VS-IMSRG effective interaction are in good agreement with the experimental data.  Further improvements are made in the IMSRG approach to develop  VS-IMSRG interaction targeted for a particular nucleus \cite{imsrg_Z_dependent}, and the calculated binding energies from this interaction for carbon to nickel isotopes agree with other large space {\it ab initio} methods.

The success of these interactions in reproducing the experimental data for $sd$-shell nuclei motivates us to study Na isotopes using these interactions. The Na isotopes have either a single or doubly open shell. The nuclear structure properties predicted by {\it ab initio} effective interactions for nuclei away from closed-(sub) shell structure are of particular interest. The nuclear systems which do not have closed-(sub) shell structures may exhibit deformation \cite{deformation_open_shell_nuclei}. Extraction of such collective properties from microscopic interactions is also challenging. The neutron-rich Na isotopes ($^{29-31}$Na) lie at the island of the inversion region and are affected by intruder configuration from higher model space. The structural properties of such isotopes obtained from {\it ab initio} effective interactions should be addressed. The energy spectra of $^{26-27}$Na were studied through $\gamma$ spectroscopy in Ref. \cite{exp_26-27Na}. The low-lying energy structures of neutron-rich $^{27-29}$Na were reported in Ref. \cite{exp_27-29Na}. In these articles, the experimental findings were compared with the shell model predictions using USDB interaction only. Recently precise measurement for the reduced $E2$-transition strength has been done for $^{23}$Na \cite{exp_23Na}. In Ref. \cite{em_properties_sd_shell}, the quadrupole and magnetic moments of Na isotopes were calculated from a mass-dependent VS-IMSRG interaction. The low-lying energy levels of only $^{22}$Na, obtained from {\it ab initio} effective interactions, are presented in Ref. \cite{sdshell_int}. The nuclear moments of neutron-rich odd mass Na isotopes were studied by Otsuka \textit{et al} in \cite{even_N_Na_moments}. 

In this article, we have done a comprehensive study of energy spectra, electromagnetic properties such as transition strengths, quadrupole moments, and magnetic moments for the Na chain in $sd$-shell ($A=$ 20 to 31) using N3LO, JISP16, DJ16A, and nucleus dependent VS-IMSRG effective interactions. The calculated results are compared with the results of the USDB interaction and available experimental data. The other nuclear observables that provide vital information regarding nuclear structure properties are root-mean-square (rms) charge and matter radii and neutron skins. The study of rms radii along the isotopic chain of nuclei can reveal information regarding nuclei's spatial extension, deformations, shape and size evolutions, and the rise of new magic numbers. The modern experiments are focusing on measurement of charge radii of different isotopic chains \cite{ch_radii_Ca_isotopes,ch_radii_proton_rich_Ca_isotopes,
ch_radii_Cu_isotopes,ch_radii_neutron_deficient_K_isotopes,ch_radii_K_isotopes,
ch_radii_Ne_Mg_isotopes,ch_radii_O_isotopes}. The charge radii are measured precisely through hyperfine structure and atomic shift measurements of nuclei. The estimation of nuclear charge radii will provide a testing ground for nuclear theory and models in stable nuclei as well as in nuclei towards the drip line. The charge and matter radii of odd-$A$ Na isotopes have been studied from effective nucleon-nucleon interactions in \cite{even_N_Na_moments,Na_atomic_and_nuclear_theory}. We have calculated the charge radii of Na isotopes with $A=$ 20 to $A=$ 31. The charge radii are evaluated using shell model harmonic oscillator (H.O.) wave functions and oscillator length parameter ($b$) in $sd$ and $sdpf$-shell. The $b$ values are taken from $\hbar\omega$ as well as from experimental charge radii data. Additionally, we have also presented a discussion on matter radii and neutron skins of the Na chain. 

The contents of this paper are organized in the following manner. We have presented a very brief discussion on the derivation of microscopic effective interactions from NCSM and IMSRG {\it ab initio} approaches in section \ref{section2}. In section \ref{section3}, we started investigating the nuclear structure properties of Na isotopes by comparing the shell model results for binding energies and energy spectra with the experimental data in subsection \ref{subsection3.1} and \ref{subsection3.2}.  The subsection \ref{subsection3.3} contains a discussion on electromagnetic properties comprising of reduced $E2$-transition strengths, quadrupole and magnetic moments of  the Na chain. The rms charge, matter radii, and neutron skins of Na isotopes are discussed in subsection \ref{subsection3.4}.

\section{Effective Interactions for $\bold{SD}$-shell} 
\label{section2}
The microscopic effective interactions, we have used to study the nuclear structure properties of Na isotopes, are N3LO, JISP16, DJ16A, and IMSRG. These effective interactions are derived for $sd$-shell using {\it ab initio} approaches. The first three interactions are derived from the NCSM formalism \cite{ab_initio_no_core_review1} while the last one through IMSRG \cite{imsrg_start} approach. A brief discussion on the method of deriving effective interactions from the NCSM and IMSRG approach is given below; however, the detailed procedure can be found in the Refs. \cite{imsrg_A_dependent,
imsrg_Z_dependent,eff_int_method(2015),sdshell_int}.

The many-body Hamiltonian for a system of $A$ nucleons can be written as
\begin{equation}\label{eq:1}
H_A=\frac{1}{A} \sum_{i<j}^{A}\frac{{(\vec{p_i}-\vec{p_j})}^2}{2m}+\sum_{i<j}^{A}V_{ij}^{NN} +\sum_{i<j<k}^{A}V_{ijk}^{NNN}.
\end{equation}
Where $m$ is the average mass of proton and neutron, the first term represents $A$ nucleon's relative kinetic energy; the second and third terms correspond to bare \textit{NN} and \textit{NNN} interactions, respectively. The Hamiltonian can not be solved directly with bare interactions since they generate strong short-range correlations. To achieve convergence in results, renormalization techniques are implemented to obtain effective interactions in the chosen model space. These effective interactions preserve all bare interactions' symmetries and reproduce the {\color{black}energy spectrum of the bare interactions} in our desired model space in the low-energy domain.
 The renormalization methods that are used in NCSM and IMSRG approaches are Okubo-Lee-Suzuki (OLS) scheme \cite{OLS1, OLS2} and Similarity Renormalization Group (SRG) \cite{SRG}, respectively.

In NCSM, all nucleons are treated as active, and we start from the Hamiltonian given in Eq. \eqref{eq:1} considering up to \textit{NN} potential. The Hamiltonian is constructed with H.O. basis states, and a center of mass H.O. Hamiltonian ($H_{cm}$) term is added (which is later subtracted) to facilitate convergence. This $H_{cm}$ term introduces $\Omega$ and $A$ dependence on the \textit{NN} potential. The modified Hamiltonian is given in Eq . \eqref{eq:2}
\begin{equation}\label{eq:2}
H_a+H_{cm}=\sum_{i=1}^a\left[\frac{{\vec{p_i}}^2}{2m}+\frac{1}{2}m{\Omega}^2\vec{r_i}^2\right] 
+\sum_{i<j=1}^{a}V_{ij}(\Omega,A).
\end{equation}
Here for $a=A$, we will get the original Hamiltonian. In the NCSM, two-body 
cluster approximation, i.e., $a$=2 is used, and the first OLS transformation is applied to derive a primary effective Hamiltonian in a space characterized by parameter N$_{max}$, $\hslash\Omega$, and $A$. Here N$_{max}$ defines the maximum no. of H.O. quanta above the $A$-nucleonic configuration, and $\hslash\Omega$ is the H.O. energy. Then a second OLS transformation is performed on this Hamiltonian and projected to $sd$-space to obtain the effective interaction. The microscopic effective interactions JISP16, N3LO, and DJ16 have been derived using NCSM formalism. In all these effective interactions, the values of single-particle energies are taken from USDB interaction \cite{usdb_interaction} as reported in \cite{sdshell_int}. The two-body matrix elements (TBMEs) for the potential N3LO, JISP16 are obtained from Ref. \cite{eff_int_method(2015)}. DJ16A is the monopole modified version of DJ16 interaction, and its TBMEs are taken from Ref. \cite{sdshell_int}.

{\color{black} Earlier in Refs. \cite{3B_clusterA3,3B_clusterA4}, the NCSM calculations have been performed at three-body cluster level for $A=3,4$ system where it was observed that the binding energy changes about 10$\%$ in going from two-body to three-body cluster level. Similar results hold for larger model spaces, such as for $A=12$ system, and it was concluded that the significant contribution to binding energy comes from two-body cluster level \cite{3B_cluster}. In a system, the effect of higher clusters can be included by considering larger model space or large $N_{max}$ value. So, in our case, the NCSM calculations were restricted at two-body cluster level with $N_{max}=4$. The inclusion of higher $N_{max}$ for our systems will require more computational resources and time \cite{eff_int_method(2015),sdshell_int}.}

In IMSRG, the SRG method is used to decouple a small, designed model space from its large complementary space by applying a continuous sequence of unitary transformations on the Hamiltonian. In practice, this is done by solving the flow equation given in Eq. \eqref{eq:3}
\begin{equation}\label{eq:3}
\frac{dH(s)}{ds}=\left[ \eta(s),H(s)\right] 
\end{equation}
where $\eta(s)$ is the anti-hermitian generator and related to the unitary transformation as $\eta(s)\equiv[\frac{dU(s)}{ds}]U^\dagger(s)=-\eta^\dagger(s)$. Here `$s$' is the so-called flow parameter. By choosing a suitable $\eta(s)$, the off-diagonal part of the Hamiltonian $H^{od}(s)$ can be driven to zero by evolving the Hamiltonian from $s\rightarrow0$ to $s\rightarrow\infty$. The evolution of the Hamiltonian may induce higher-order forces which are truncated because of computational limit \cite{imsrg_review_H_Hergret}.
The  IMSRG calculations were performed starting from a chiral \textit{NN} interaction at N3LO, and a chiral \textit{NNN} interaction at N2LO evolved through SRG in H.O. basis in Ref. \cite{imsrg_Z_dependent}.  After that, it is transformed into the Hartree-Fock basis state, and normal ordering is done w.r.t an ensemble reference state. The normal ordered 0-, 1-, and 2-body parts of the Hamiltonian (Eq. \eqref{eq:4}) contains the in-medium contribution of \textit{NNN} forces. 
\begin{equation}\label{eq:4}
H=E_0+\sum_{ij}f_{ij}{a^\dagger_ia_j}+\frac{1}{4}\sum_{ijkl}\Gamma_{ijkl}{a^\dagger_ia^\dagger_ja_la_k}
\end{equation}
 Then a valence space interaction is decoupled from the large Hilbert space using a suitable form of $\eta(s)$. The single-particle energies and TBMEs of the IMSRG effective interaction are taken from the Ref. \cite{imsrg_Z_dependent}. This interaction is targeted for a particular nucleus. {\color{black}However, later an error is reported in the fitting procedure of \textit{NNN} interactions used in Ref. \cite{imsrg_Z_dependent}. One can find more details  in Refs. \cite{3nerror1,3nerror2}.}

In addition to these microscopic effective interactions, we have also used phenomenological USDB interaction \cite{usdb_interaction}. 
We have performed the shell model calculation for Na isotopes using KSHELL code \cite{Kshellcode}. The ground state energies relative to the core $^{16}$O for sd-shell nuclei are determined from Eq. \eqref{eq:5}
\begin{equation}\label{eq:5}
	E(A,Z)^r = E(A,Z) - E(^{16}{\rm O})-E_c(Z).
\end{equation}
Here $E(A,Z)^r$ and $E(A,Z)$ are the relative and absolute ground state energies for the nucleus of mass number $A$ and atomic no. $Z$. The second and third terms of Eq. \eqref{eq:5} are the binding energy of the core $^{16}$O and Coulombic correction energy (for $Z$=11), which have the values -127.619 MeV \cite{NNDC} and 11.73 MeV \cite{Coulomb_corr_energy}, respectively.
\section{Results and discussion} \label{section3}
\subsection{{\bf Ground state Energy}}\label{subsection3.1}
\begin{figure}
	\includegraphics[width=95mm]{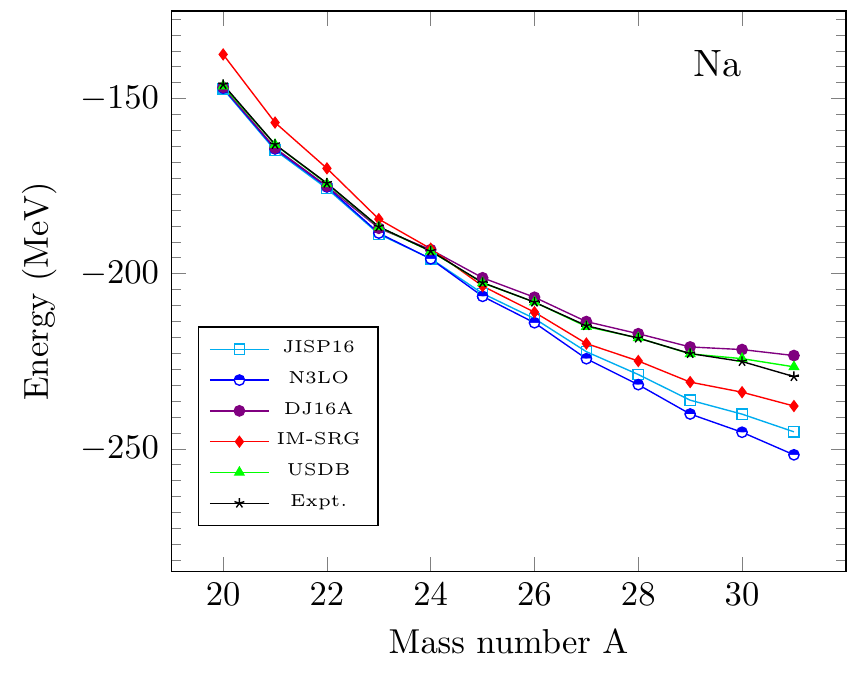}
	\centering
	\caption{Comparison of calculated ground state energy of Na isotopes with the experimental data \cite{NNDC}.}
	\label{fig:1}
\end{figure}

The g.s. energies have been calculated for the Na chain using Eq. \eqref{eq:5}. Fig. \ref{fig:1} shows the g.s. energies of Na isotopes calculated from all microscopic effective interactions as well as from the phenomenological USDB interaction. The USDB interaction has excellent agreement with experimental data, followed by the DJ16A interaction. The energies obtained from JISP16 and N3LO interaction are also in good agreement with the experimental data, but as the number of valence nucleons increases, they start overbinding the g.s. energy. In the higher mass region of Na isotopes, the IMSRG results also overbind the g.s energy, but its results are better than JISP16 and N3LO. The g.s. binding energies of neutron-rich Na isotopes are not well reproduced by these interactions. However, we find a good agreement of  DJ16A interaction with experimental values. The results of the DJ16A interaction slightly underbind the g.s energies in the neutron-rich isotopes compared to other microscopic effective interactions.
\subsection{{\bf Energy spectra}}\label{subsection3.2}
\begin{figure}
	\includegraphics[width=70mm,height=55mm]{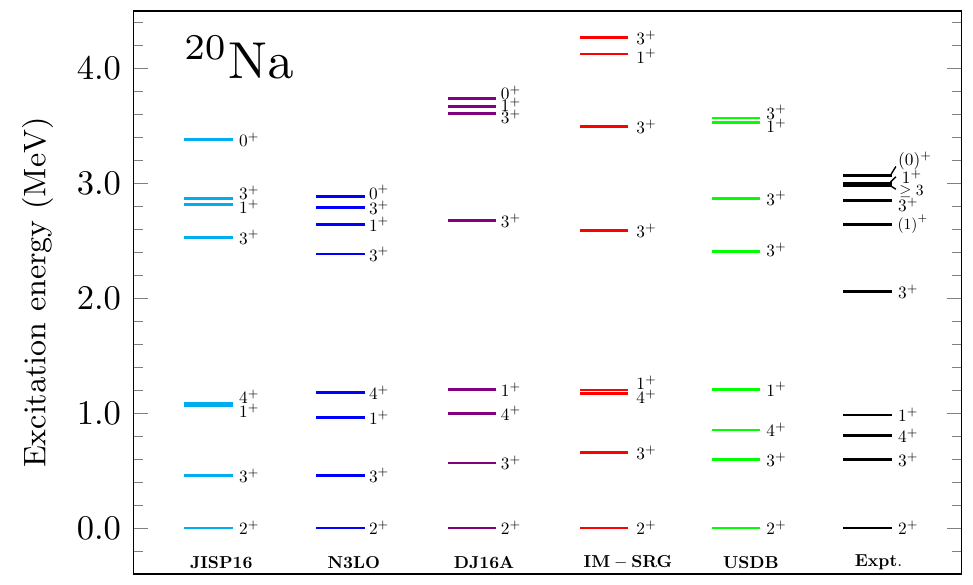}
	\includegraphics[width=70mm,height=55mm]{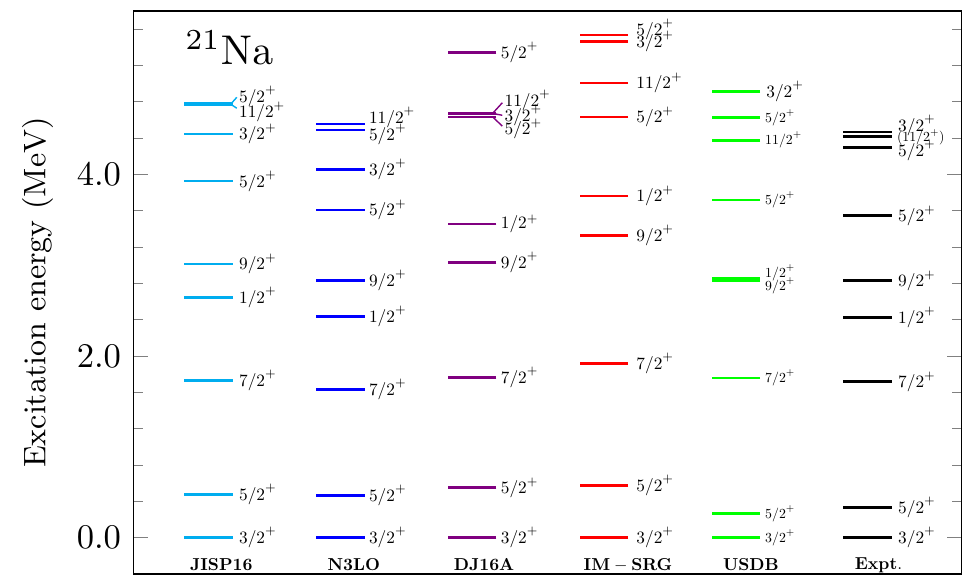}
	\includegraphics[width=70mm,height=55mm]{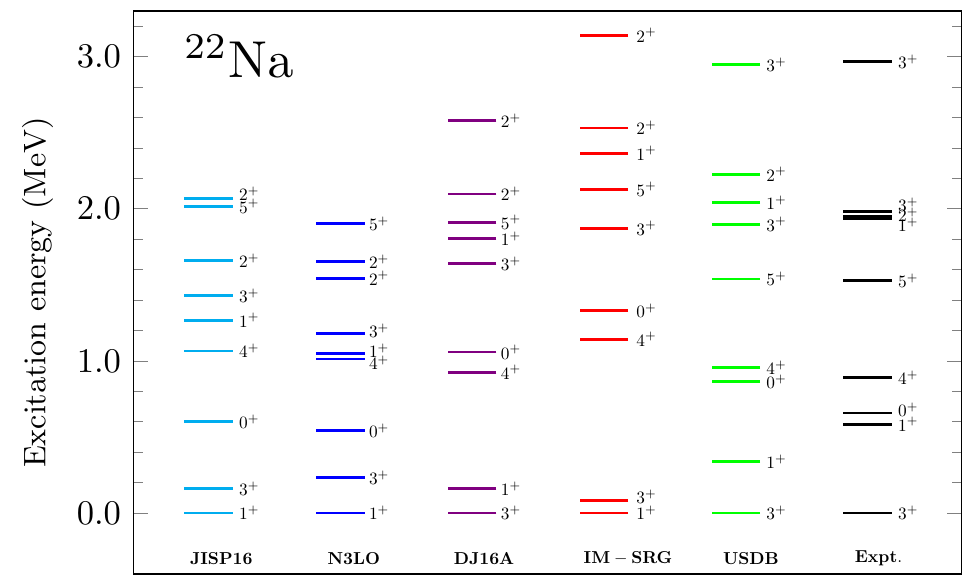}
	\includegraphics[width=70mm,height=55mm]{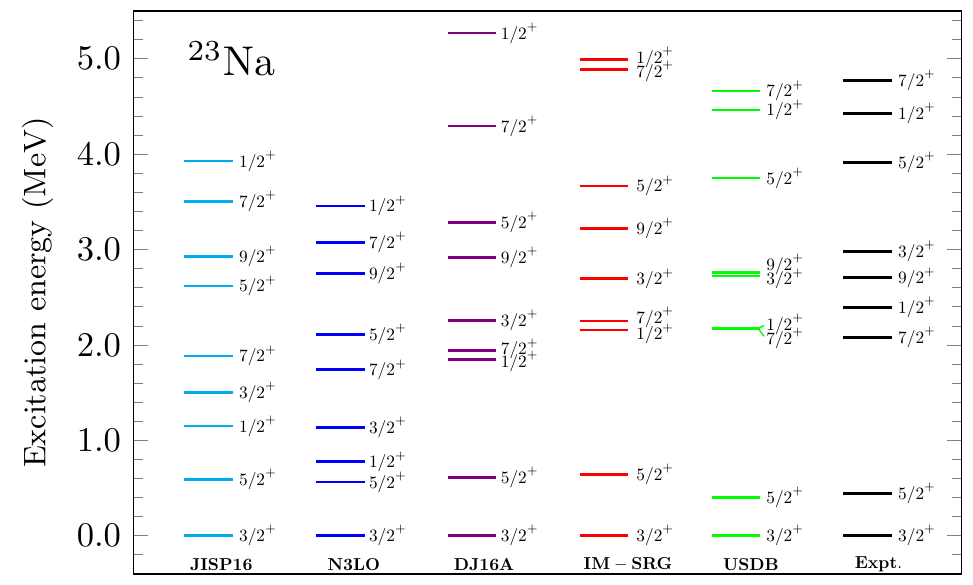}
	\includegraphics[width=70mm,height=55mm]{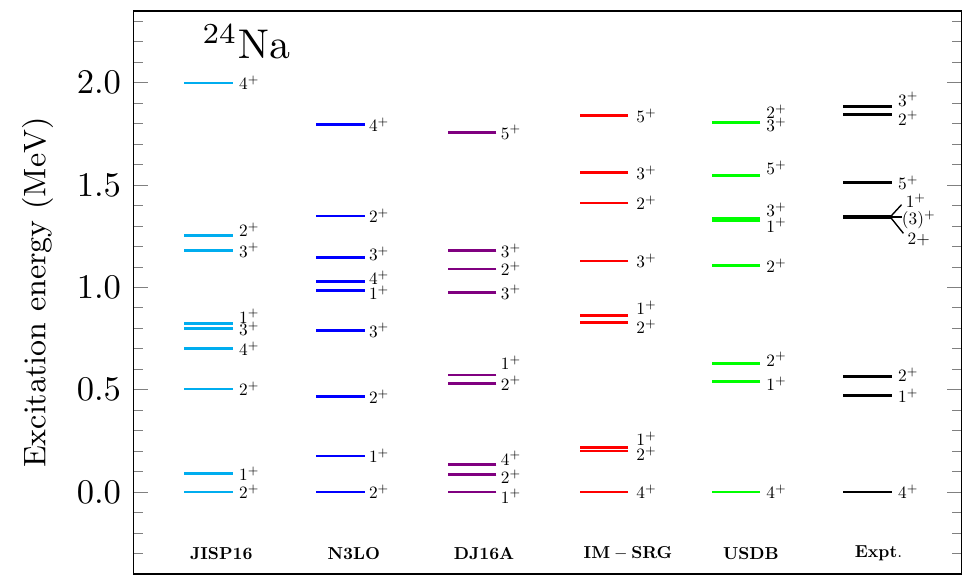}
	\includegraphics[width=70mm,height=55mm]{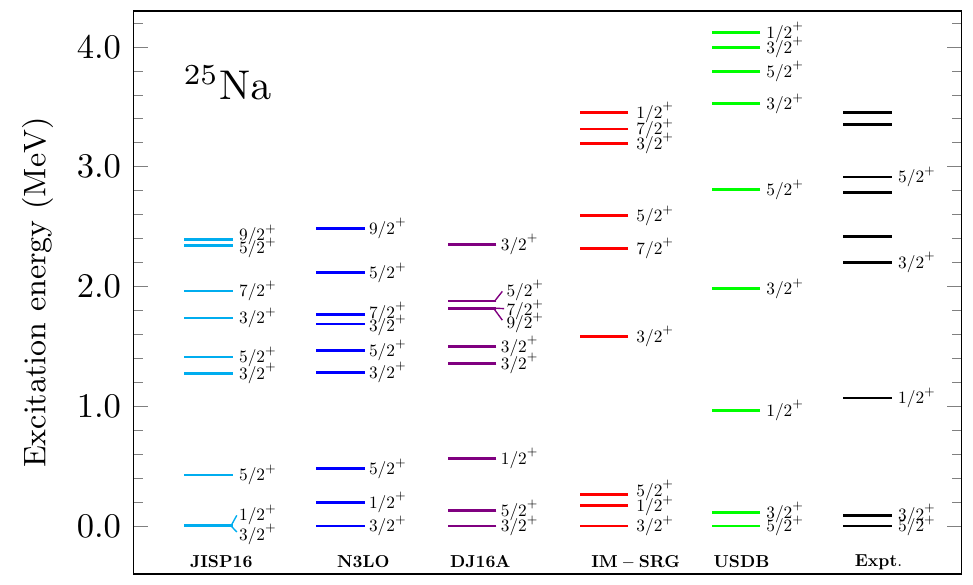}	
	\caption{Comparison between calculated and experimental \cite{NNDC} energy levels for $^{20-25}$Na.}
	\label{fig2}
\end{figure}

\begin{figure*}
	\includegraphics[width=70mm,height=55mm]{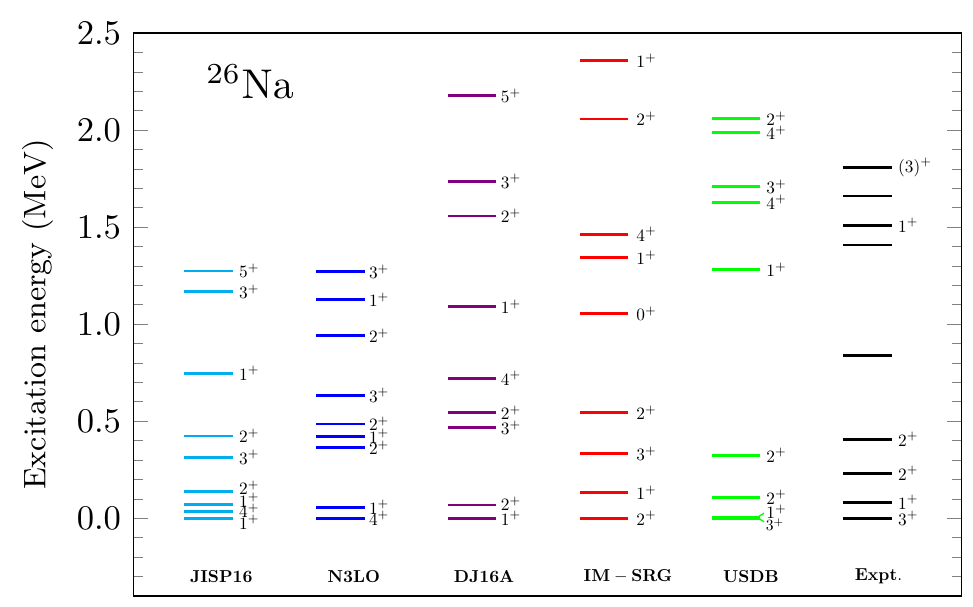}
	\includegraphics[width=70mm,height=55mm]{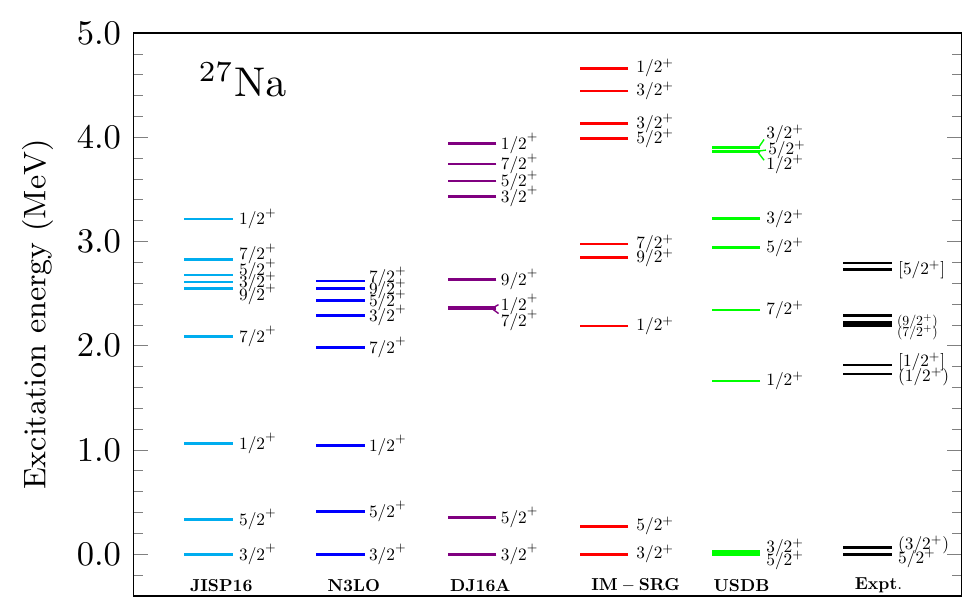}
	\includegraphics[width=70mm,height=55mm]{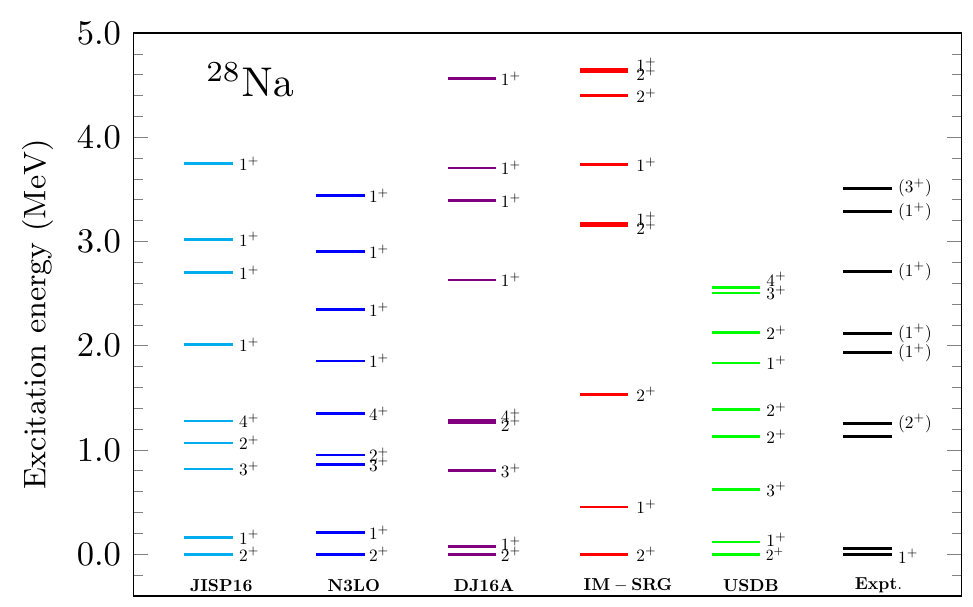}
	\includegraphics[width=70mm,height=55mm]{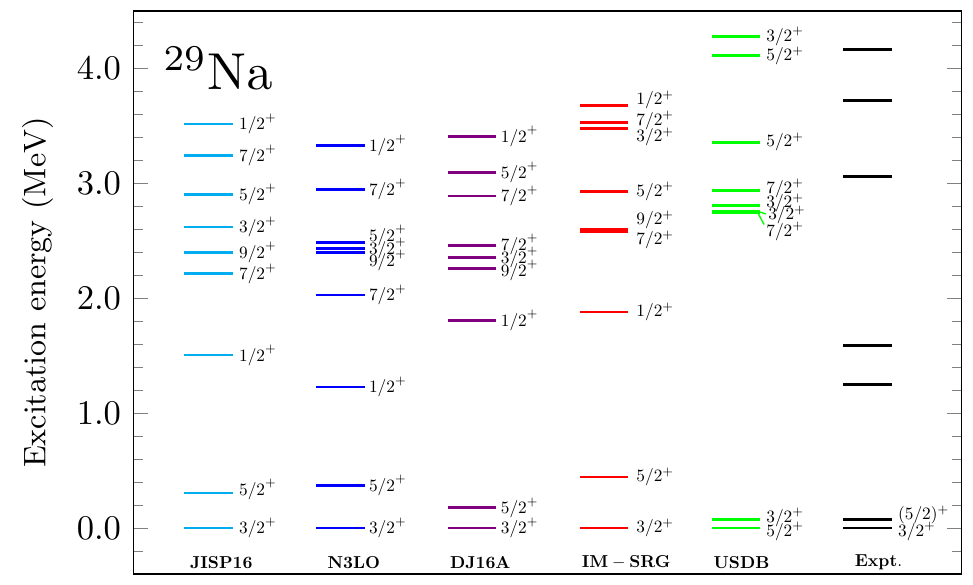}
	\includegraphics[width=70mm,height=55mm]{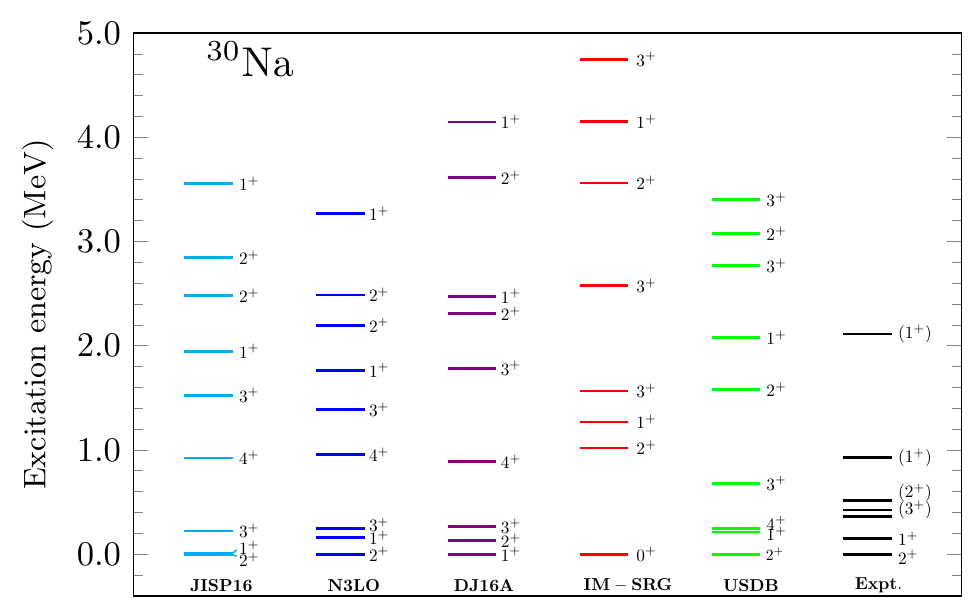}
	\includegraphics[width=70mm,height=55mm]{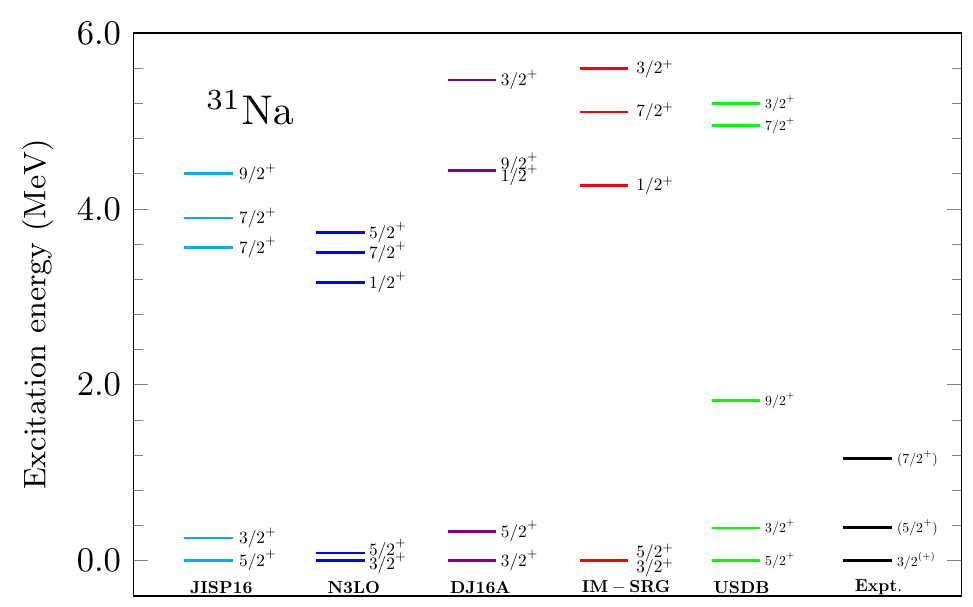}	
	\caption{Comparison between calculated and experimental \cite{NNDC} energy levels for $^{26-31}$Na.}
	\label{fig3}
\end{figure*}

The low-energy spectrum of Na isotopes is shown in Fig. \ref{fig2} and Fig.  \ref{fig3}. The microscopic effective interactions are quite good in reproducing the experimental energy spectra in lighter Na isotopes. 
The DJ16A interaction correctly reproduces the g.s. spin in $^{20-23}$Na. The other microscopic effective interactions also reproduce the g.s. spin correctly in these isotopes, except for $^{22}$Na. The g.s. (2$^+$) wave function of $^{20}$Na has a configuration of $\ket{\pi (d_{5/2}^3) \otimes \nu (d_{5/2}^1)}$ with a probability 41.77 $\%$ with USDB interaction. All effective interactions also show a dominant probability for this configuration. In $^{20}$Na, the energy of the first excited state $3^+_1$ from IMSRG interaction is slightly higher than the experimental value. In contrast, the same state appears at a bit lower energy in the energy spectra obtained from JISP16 and N3LO interaction. This $3^+_1$ state lies very close to the experimental level with a difference of only 31 keV with DJ16A interaction. There is an inversion in the lowest $4^+$and $1^+$ states in the spectra calculated from JISP16 and N3LO interaction compared to the observed data. The IMSRG interaction predicts the correct order of these states, but they lie very close to each other as compared to the experimental result. These states are shifted to slightly higher energies in the spectra of DJ16A interaction; still, the energy difference between them is in close agreement with the experimental value and better than that obtained from USDB interaction. The experiment predicts an unconfirmed energy level between the $3^+_2$ and $3^+_3$ states and suggests its spin as 1. The IMSRG, DJ16A, and USDB interactions fail to reproduce this energy level. However, both JISP16 and N3LO interactions predict the same spin, i.e., 1, as suggested by the experiment.

 In all interactions, the g.s. (3/2$^+$) of $^{21}$Na results from large mixing of configurations. The first and second excited states are well reproduced by all interactions in $^{21}$Na. The lowest 1/2$^+$ and 9/2$^+$ states are in the same order, as found in the experiment, with N3LO and JISP16 interactions.
We obtained a reverse order for these states from USDB and other microscopic effective interactions. The USDB and N3LO interactions predict 11/2$^+$ and 3/2$^+$ states between 5/2$^+_2$ and 5/2$^+_3$, respectively,  whereas both 11/2$^+$ and 3/2$^+$ states occur between these states in the spectra of DJ16A, JISP16, and IMSRG interactions. However, the experiment predicts no such states between 5/2$^+_2$ and 5/2$^+_3$. The overall spectra calculated from N3LO interaction for $^{21}$Na are in better agreement with the experimental data.

The $\ket{\pi (d_{5/2}^3) \otimes \nu (d_{5/2}^3)}$ configuration has a probability of 23.82$\%$ in g.s. (3$^+$) of $^{22}$Na with USDB interaction, while in DJ16A this configuration contributes only 19.2$\%$ for the g.s. Except for DJ16A, all microscopic effective interactions predict $1^+$ state as their g.s. The experiment predicts the first excited state 1$^+$ at 583 keV.  However, this state appears at an energy 338 and 162 keV in USDB and DJ16A interactions, respectively. In addition to USDB, only JISP16 and N3LO interactions predict the same order of the lowest 0$^+$ and 4$^+$ states as found in experiment. 

In $^{23}$Na, the first excited state (5/2$^+$) calculated from all microscopic effective interactions is at relatively higher energies than the observed value. However, this state appears at an energy lower than 41 keV from the experimental value in USDB spectra. The N3LO and JISP16 interactions poorly reproduce the observed energy spectra of $^{23}$Na. Contrary to experimental spectra, the 7/2$^+_1$ and 1/2$^+_1$ states are inverted in the DJ16A and IMSRG spectra. With these interactions, the 9/2$^+_1$ and 3/2$^+_2$, 1/2$^+_2$, and 7/2$^+_2$ states are also inverted. Additionally, the USDB interaction also fails to predict the correct order of the states 9/2$^+_1$ and  3/2$^+_2$. In $^{23}$Na, DJ16A and IMSRG interactions predictions are better than N3LO and JISP16 interactions.

For the medium mass Na isotopes, the predictions of the microscopic effective interactions for energy spectra are not as good as in the case of lighter Na isotopes. In the energy spectra of $^{24}$Na, only USDB and IMSRG interactions reproduce the experimental g.s. 4$^+$. Both of these interactions predict the g.s. configuration as $\ket{\pi (d_{5/2}^3) \otimes \nu (d_{5/2}^5)}$ with probabilities 31.4$\%$ and 22.7$\%$ respectively. The energy spectra in $^{25}$Na and $^{26}$Na are not well described by any microscopic effective interactions. In $^{27}$Na, the observed ground state appears as the first excited state in the energy spectra calculated from all microscopic effective interactions. The experiment predicts the ground state of $^{28}$Na as 1$^+$; however, in the energy spectra obtained from all interactions, 1$^+$ state appears above the ground state. The energy shift of the 1$^+$ state relative to g.s. is highest in IMSRG, followed by  N3LO and JISP16 interactions. The energy shift of this 1$^+$ state calculated from DJ16A interaction is only 77 keV which is better than that calculated from USDB interaction (117 keV).

As we move to higher mass Na isotopes towards the nuclei lying at the edge of the $N=20$ shell gap, the microscopic effective interactions are pretty good in reproducing the low-lying spectra. It means even without including $pf$- shell in the model space, microscopic interactions are able to reproduce low-lying energy states. This might be due to initially {\it ab initio} interactions are taken for multi $\hbar \Omega$ before projecting for $sd$- model space.

The  g.s. spin in $^{29}$Na is well reproduced by all microscopic effective interactions, while USDB interaction fails to reproduce it. The N3LO interaction predicts the g.s. configuration as {\color{black} $\ket{\pi (d_{5/2}^3) \otimes \nu (d_{3/2}^2 d_{5/2}^6 s_{1/2}^2)}$} with probability 45.11$\%$ . The other microscopic effective interactions predict the same configuration for the g.s. with probabilities more than 50$\%$. 
Experimental data are unavailable on the spin-parities of excited states in $^{29}$Na. The experiment suggests the spin of the first excited state to be 5/2$^+$. All of the microscopic effective interactions agree with it. The second excited state appears at energy 1249 keV relative to g.s. in the experimental energy spectra. All the microscopic effective interactions predict the spin of this state as 1/2$^+$. This 1/2$^+$ state is present at higher energies than the observed value in IMSRG spectra, followed by DJ16A and JISP16. But this state appears at 1228 keV energy in N3LO energy spectra, which is very close to the experimental value. The results of the microscopic effective interactions are better than the USDB interaction. Our calculated results for several excited states might be useful to compare future experimental data.

In the energy spectra of $^{30}$Na, the IMSRG interaction fails to reproduce the correct g.s. spin. The observed g.s. 2$^+_1$ and first excited state 1$^+_1$ are reversed in spectra obtained from DJ16A interaction. But N3LO and JISP16 interactions show the correct order of these states, as observed in the experiment. In these interactions, the g.s. arises from a configuration of {\color{black} $\ket{\pi (d_{5/2}^3) \otimes \nu (d_{3/2}^3 d_{5/2}^6 s_{1/2}^2)}$} with probabilities 69.56$\%$ and 74.36$\%$, respectively. The USDB interaction also predicts the same order and configuration for g.s. with probability 84.67$\%$. In the JISP16 spectra, the 1$^+_1$ state is very close to the ground state, i.e., it lies within 9 keV energy difference, while the experiment shows this state at an energy of 150 keV relative to the ground state. In the USDB interaction, the 1$^+_1$ state lies at higher energy than the experimental value.
But N3LO interaction predicts it at 157 keV above g.s., which is nearly at the same level as found in experimental data.

The experimentally observed  g.s. of $^{31}$Na is 3/2$^{(+)}$. But its parity is unconfirmed.  The results obtained from N3LO, IMSRG, and DJ16A interactions show the g.s. spin as 3/2$^+$, supporting the experimental data. In all these three interactions, the dominant configuration for the ground state is {\color{black} $\ket{\pi (d_{5/2}^3) \otimes \nu (d_{3/2}^4 d_{5/2}^6 s_{1/2}^2)}$}. In the experimental spectra, the first excited state is observed at 375 keV relative to the ground state. The unconfirmed spin-parity of this state is 5/2$^+$. Our calculated spectra from all microscopic effective interactions, except JISP16, predict the same spin-parity for the first excited state. In IMSRG spectra, the g.s. and the first excited state lie almost at the same energy. The energy difference between these states is slightly higher in N3LO spectra than IMSRG spectra but still less than the experimental value. The 5/2$^+$ state appears at 332 keV energy with DJ16A, which agrees reasonably with the experimental data. The USDB and JISP16 interactions show a reverse order of the observed ground and first excited states. The g.s. and first excited states in $^{31}$Na are well described by DJ16A interaction than others.

For low-mass Na isotopes, i.e.,  in the case of $^{20,21,23}$Na, the microscopic effective interactions are quite good in reproducing the g.s. energy and low-lying states because of a smaller number of valence particles in the model space. But as the number of neutrons increases in the valence space, the correlation energy among these particles increases. 
A comparative study for monopole modification on the microscopic effective interaction has been done in Ref. \cite{sdshell_int}. It is found that the DJ16 interaction needs small modifications than others in order to improve the agreement of the calculated results with the experimental data. 
For $A=$ 24 and above, the g.s. energy and the energy spectra obtained from DJ16A interaction are in better agreement with experimental data than other microscopic effective interactions. 
The calculated results of $^{22}$Na from all microscopic effective interactions except for DJ16A are deviating from the experimental data, thus as mentioned in Ref. \cite{22Na_3N} \textit{NNN} forces are important for describing the energy spectra for such a nucleus.
For the Na isotopes, as we approach towards the island of inversion region ($^{29-31}$Na), the predictions of the microscopic effective interactions for the spin-parities of the g.s and first excited states are better than the USDB interaction. 

In Ref. \cite{eff_int_method(2015)}, the effective interactions were developed for $sd$-space from original N3LO and JISP16 interactions by performing NCSM calculation on $^{18}$F. The same procedure is followed in Ref. \cite{sdshell_int} to develop DJ16 interaction. Since the calculations were done with  $^{18}$F, the residual Coulombic interactions between the valence protons are ignored. But in Na isotopes, three active protons exist in the valence space. So an effective interaction, which is developed by considering the residual Coulombic interactions, may improve the calculated energy spectra of Na isotopes. The effective interactions N3LO, JISP16, and DJ16A do not retain the complete charge dependency. The procedure for maintaining the total charge dependency is described in Ref. \cite{eff_int_method(2015)}, and a study on this aspect is also in progress by the same group. The valence space interaction that includes complete charge dependency may provide more precise results on the spectroscopy of Na isotopes. To further increase the predictive power of these microscopic effective interactions, it is desirable to include a larger N$_{max}$ in the calculation while obtaining the effective interactions for $sd$ model space. Additionally, the energy spectra of the neutron-rich Na isotopes ($N=$10,11,12) may be better understood through effective interaction projected to the $sd-pf$ shell.
Recently,  Ne, Mg, and Si isotopes in the island of the inversion region have been well described with VS-IMSRG interaction developed for multi-major shells, i.e., spanning $sd-pf$ model space \cite{imsrg_sd_pf}.  
\subsection{{\bf Electromagnetic properties}}\label{subsection3.3}
In this section, we have discussed the shell model results for reduced electric quadrupole transition strengths, quadrupole, and magnetic moments of the Na chain. {\color{black}The validity of the effective $E2$ and $M1$ operators derived through the consistent IMSRG and OLS transformations has been studied in $p$-, $sd$- and $fp$-shell nuclei \cite{EffOperatorE2M1_1,EffOperatorE2M1_2}. As the description of $E2$ strengths in $sd$-shell by the effective operator, in particular, is found to be extremely poor \cite{EffOperatorE2M1_2}, we here employ phenomenological effective charges for the study of quadrupole observables. 
We also do not use the effective M1 operator since the effects of consistent evolution of the M1 operator results in rather mild modifications to the bare M1 matrix elements, which still needs to be improved with the inclusion of the meson-exchange current effects \cite{EffOperatorE2M1_2}.}
The shell model calculations are performed with effective charges e$_{\pi}$=1.5e and e$_{\nu}$=0.5e and free $g$-factors ($g^{free}_l$ and $g^{free}_s$).

The $E2$ strengths for transitions between excited states and g.s or between excited states are essential in studying collective properties of the nucleus arising due to deviation from spherical symmetry. The microscopic effective interactions have given good results for the g.s. energies and low-lying spectra. In this section, we have further used these microscopic effective interactions to calculate the $E2$ strengths for selected transitions in Na isotopes. The results of these microscopic effective interactions and USDB interactions are then compared with the corresponding experimental data \cite{exp_23Na,NNDC,BE2_29Na}.  Table \ref{tab:BE2_values} contains the $B(E2)$ values of Na isotopes calculated for various transitions.

In $^{21}$Na, the $E2$ strength for the transition ${5/2}_{1}^{+}\rightarrow {3/2}_{1}^{+}$ obtained from DJ16A interaction is  better than all other interactions including USDB. All interactions predict larger $B(E2)$ values for the transition ${7/2}_{1}^{+} \rightarrow {5/2}_{1}^{+}$ compared to the experimental data. The IMSRG and USDB results for this transition are very close, while the values obtained from DJ16A, JISP16, and N3LO are larger than the USDB result.
All microscopic effective interactions predict weak transition strengths for ${7/2}_{1}^{+} \rightarrow {3/2}_{1}^{+}$, but still, their results are better than the result of USDB interaction.

The USDB result is close to the experimental $B(E2)$ value for $1^+_1 \rightarrow 3^+_1$ transition in $^{22}$Na while IMSRG interaction predicts a lesser $B(E2)$ value for it. All other effective interactions produce large $B(E2)$ values for this transition. For the transition $1^+_2 \rightarrow 3^+_1$, the experimental $E2$ strength is 9.5(32) $ e^2fm^4$. The N3LO interaction correctly reproduces this value, while other interactions, including USDB, predict a larger $B(E2)$ value for this transition. We have also obtained larger $B(E2)$ values from all effective interactions for the transition $5^+_1 \rightarrow 3^+_1$ compared to the observed $B(E2)$ value. The USDB and all microscopic effective interactions poorly reproduce the transition strengths in transitions $4^+_1 \rightarrow 3^+_1$,$2^+_1 \rightarrow 0^+_1$ and $3^+_2 \rightarrow 1^+_1$.
\begin{table*}
	\centering
	\caption{The electromagnetic transition rates ($B(E2)$ values) of Na isotopes in $e^2fm^4$ unit. }
		\begin{tabular}{ccccccccc}
		\hline 
	Nuclei & A	&	$J_i^{\pi} \rightarrow J_f^{\pi}$	&	N3LO	&	JISP16	&	DJ16A   &   IMSRG  &	USDB&	Exp. \cite{NNDC}	\T\B \\
	\hline
	
Na  &21	&	${5/2}_{1}^{+}\rightarrow{3/2}_{1}^{+}$  & 112.4     & 114.6 & 119.2 & 111.9  &  109.9 &  134(10)\T\B \\
	&		&	${7/2}_{1}^{+}\rightarrow{5/2}_{1}^{+}$  & 86.3     & 85.0 & 81.5&  73.7   &  72.3 &  55(27)\T\B \\
	&		&	${7/2}_{1}^{+}\rightarrow{3/2}_{1}^{+}$  & 51.7     & 50.9 & 49.9& 46.9    &  45.8 &  72(27)\T\B \\

	&22		&	$1_{1}^{+}\rightarrow3_{1}^{+}$  &8.1      & 7.1 & 0.2& 0.002  &  0.015 &  0.034(2)\T\B \\
	&		&	$4_{1}^{+}\rightarrow 3_{1}^{+}$  & 129.5     & 127.7 & 126.2&  117.7   &  112.7 &  91.9(32)\T\B \\
	&		&	$5_{1}^{+}\rightarrow3_{1}^{+}$  & 29.4     & 28.6 & 27.5&    27.1 &  25.9 &  19(1)\T\B \\		
	&		&	$1_{2}^{+}\rightarrow3_{1}^{+}$  & 9.4     & 12.9 & 14.8 &    15.2 &  16.3 &  9.5(32)\T\B \\
	&		&	$2_{1}^{+}\rightarrow0_{1}^{+}$  & 77.8     & 77.0 & 75.8 &    71.9 &  69.6 &  55(18)\T\B \\
	&		&	$3_{2}^{+}\rightarrow1_{1}^{+}$  & 85.6     & 88.2 & 93.0 &     89.5 &  86.6 &  69.2(73)\T\B \\

  &23	&	${5/2}_{1}^{+}\rightarrow{3/2}_{1}^{+}$  & 135.6     & 140.4 & 139.4&  136.4   &  133.1 &  106(4) \cite{exp_23Na} \T\B \\
	&		&	${7/2}_{1}^{+}\rightarrow{5/2}_{1}^{+}$  & 98.4     & 93.7 & 86.4 &  76.2   &  69.7 &  56.7(85)\T\B \\
	&		&	${7/2}_{1}^{+}\rightarrow{3/2}_{1}^{+}$  & 57.7     & 55.2 & 54.0  &  49.5   &  46.7 &  47.4(58)\T\B \\
	&		&	${1/2}_{1}^{+}\rightarrow{5/2}_{1}^{+}$  & 0.1     & 2.3 & 5.7  &  11.7 &  17.7 & 11.2(27)\T\B \\		
		&		&	${9/2}_{1}^{+}\rightarrow{5/2}_{1}^{+}$  & 73.4     & 74.3 & 73.8 &  71.9   &  69.5 &  67.6(54)\T\B \\
			&		&	${3/2}_{2}^{+}\rightarrow{5/2}_{1}^{+}$  & 8.4     & 6.7 & 7.8  &  5.4   &  3.6 &  5.4(38)\T\B \\

&24		&	$(3)_{1}^{+}\rightarrow4_{1}^{+}$  & 0.5     & 0.9 & 0.4 & 1.1    &  0.7 & NA \T\B \\
	&		&	$2_{3}^{+}\rightarrow1_{1}^{+}$  & 10.7     & 2.8 & 0.02  & 2.5    &  1.7 &  4.9(41)\T\B \\
		&		&	$3_{2}^{+}\rightarrow4_{1}^{+}$  & 0.6     & 0.4 & 0.008 &  0.3   &  1.6 &  1.6(10)\T\B \\

  &25	&	${3/2}_{1}^{+}\rightarrow{5/2}_{1}^{+}$  & 195.6     & 173.4 & 165.03&  148.9   &  155.6 & NA  \T\B \\ 
	&		&	${1/2}_{1}^{+}\rightarrow{5/2}_{1}^{+}$  & 15.3     & 40.0 & 19.2   &  77.8   &  51.0 &  30.8(47)\T\B \\

&26		&	$1_{1}^{+}\rightarrow3_{1}^{+}$  & 57.0  & 74.0 & 54.3  &  38.4   &  42.7 &16.5(41)  \T\B \\
		&		&$2_{1}^{+}\rightarrow3_{1}^{+}$  & 34.8     & 78.3  & 67.6  &  88.6   &  25.4 &  50(45)\T\B \\

&29		&	${(5/2)}_{1}^{+} \rightarrow {3/2}_{1}^{+}$  & 107.42  & 107.65 & 106.37  &  98.87   &  95.88 & 93(16) \cite{BE2_29Na} \T\B \\

&30		&	$1_{1}^{+} \rightarrow 2_{1}^{+}$  & 1.16  & 0.36 & 0.02  &  0.9   &  0.06 & NA  \T\B \\

&31		&	${(5/2)}_{1}^{+} \rightarrow {3/2}_{1}^{+}$  & 78.60  & 77.24 & 75.27  &  76.39   &  75.91 & NA  \T\B \\

\hline 

\end{tabular}
\label{tab:BE2_values}
\end{table*}

One can see from the Table \ref{tab:BE2_values} that all effective interactions show an enhanced $B(E2)$ value for the transitions ${5/2}_{1}^{+} \rightarrow {3/2}_{1}^{+}$ and ${7/2}_{1}^{+} \rightarrow {5/2}_{1}^{+}$ in $^{23}$Na. The measured transition strength between the lowest 7/2$^+$ and g.s. 3/2$^+$ is 47.4(58) $ e^2fm^4$. The USDB interaction predicts a slightly smaller $B(E2)$ value than the experimental value for this transition, while the calculated $B(E2)$ values from DJ16A, JISP16 and N3LO are comparatively larger than the experimental data. But, from IMSRG interaction, we found $B(E2; {7/2}_{1}^{+} \rightarrow {3/2}_{1}^{+}$) = 49.5 $ e^2fm^4$ which is very close to the observed value. The $B(E2)$ values for the transition ${1/2}_{1}^{+} \rightarrow {5/2}_{1}^{+}$ obtained from N3LO, JISP16 and DJ16A interactions are significantly weaker than the  measured value. The USDB interaction also predicts a slightly larger $B(E2)$ value for this transition, but the $B(E2)$ value obtained from the IMSRG interaction perfectly agrees with the experimental value. The $E2$-transition strength for ${3/2}_{2}^{+} \rightarrow {5/2}_{1}^{+}$ obtained from both experiment and  IMSRG interaction is 5.4 $ e^2fm^4$. The N3LO, DJ16A, and JISP16 predict a larger $B(E2)$ value for this transition, respectively, while the $B(E2)$ value from the USDB interaction is smaller than the experimental value. The $B(E2)$ values calculated for the transitions ${9/2}_{1}^{+} \rightarrow {5/2}_{1}^{+}$ from all microscopic effective interactions are in good agreement with the experimental data and slightly larger than it.

No experimental data available for $B(E2; (3)_{1}^{+} \rightarrow 4_{1}^{+}$) in $^{24}$Na. We have presented the calculated $B(E2)$ values for this transition obtained from all microscopic effective interactions and USDB interaction in Table \ref{tab:BE2_values}. The $B(E2; 2_{3}^{+} \rightarrow 1_{1}^{+}$) obtained from the N3LO interaction is much larger than the observed value. In contrast, the values obtained from other interactions, including USDB, are relatively weak compared to experimental data. The $E2$ strength for the transition $3^+_2 \rightarrow 4^+_1$ is well defined by USDB interaction. All other microscopic effective interactions predict weak $B(E2)$ values for this transition.

The $E2$ strength for the transition from the  3/2$^+_1$ to g.s. 5/2$^+_1$ is not known experimentally in $^{25}$Na. We observed that the $B(E2; {3/2}_{1}^{+} \rightarrow {5/2}_{1}^{+}$) values from DJ16A, JISP16, and N3LO interactions are larger than that obtained from USDB interaction while the $B(E2)$ value calculated from IMSRG interaction is close to USDB result and smaller than it. The experimental $E2$-transition strengths in $^{25}$Na and $^{26}$Na are not well reproduced by any of the effective interactions.
We have also calculated the $B(E2)$ values for transition between the g.s. and the first excited state of neutron-rich Na isotopes ($^{29-31}$Na). The calculated $E2$ strengths for $^{29}$Na from all effective interactions are in reasonable agreement with the experimental value. Theoretically, a very weak $E2$ strength is found in the case of $^{30}$Na.
From the detailed analysis of electric quadrupole transition strengths above, one can see that the microscopic interactions and 
\begin{landscape}
\begin{table*}
\centering
\caption{{\bf Quadrupole and magnetic moments of Na isotopes.}}
\begin{tabular}{lccccccccccccccc}
\hline
&                     &     & \multicolumn{5}{c}{$Q$ (eb)}    & \multicolumn{5}{c}{$\mu$ ($\mu_N$)} \T\B\\
\cline{4-9}
\cline{10-15}

Nuclei  & $A$  & J$^{\pi}$ & N3LO& JISP16 & DJ16A & IMSRG & USDB & Exp. \cite{IAEA}
& N3LO & JISP16 & DJ16A & IMSRG & USDB  & Exp. \cite{IAEA}\T\B\\
\hline

Na      & 20  & 2$^+$  & 0.123 & 0.120  & 0.115  & 0.101  & 0.095 & 0.101(8)
& 0.715  & 0.796  & 0.716 & 0.441 & 0.446 & 0.3694(2)\T\B\\

& 21 &3/2$^+$& 0.129  & 0.128 & 0.126 & 0.122  & 0.122  & 0.138(11)
& 2.654 & 2.645  & 2.589 & 2.436 & 2.489 & 2.38630(10)  \T\B\\

&  &5/2$^+$& -0.067  & -0.062    & -0.054 & -0.05  & -0.047 & NA
& 3.508 & 3.439  & 3.177 & 3.147 & 3.355 & 3.7(3)  \T\B\\

& 22 &3$^+$& 0.251  & 0.258    & 0.259 & 0.251  & 0.251 & 0.180(11)
& 1.816 & 1.811  & 1.792 & 1.807 & 1.790 & 1.746(3)  \T\B\\

&  &1$^+$& -0.107  & -0.117  & -0.124 & -0.123  & -0.123 & NA
& 0.654 & 0.610  & 0.545 & 0.505 & 0.517 & 0.535(10)  \T\B\\

& 23 &3/2$^+$& 0.135  & 0.132    & 0.129 & 0.123  & 0.118 & 0.104(1)
& 2.297 & 2.194  & 2.209 & 2.052 & 2.098 & 2.217522(2)  \T\B\\

& 24 &4$^+$& 0.314  & 0.314 & 0.304 & 0.299  & 0.281 & NA 
& 1.559 & 1.512  & 1.511 & 1.523 & 1.631 & 1.6903(8)  \T\B\\

&  &1$^+$& 0.049  & 0.051 & 0.052 & 0.043  & 0.052 & NA
& 0.718 & 0.896  & 0.828 & -1.409 & -1.866 & -1.931(3)  \T\B\\

& 25 &5/2$^+$& -0.027  & -0.002  & 0.005 & 0.016  & 0.002 & 0.0015(3)
& 2.872 & 2.912  & 2.945 & 2.918 & 3.367 & 3.683(4)  \T\B\\

& 26 &3$^+$& 0.010  & 0.015    & -0.009 & 0.0003  & -0.005 & -0.0053(2)
& 1.573 & 1.949  & 2.566 & 2.499 & 2.631 & 2.851(2)  \T\B\\

& 27 &5/2$^+$& -0.023  & -0.019 & -0.016 & -0.007  & -0.013 & -0.0071(3)
& 3.180 & 3.283  & 3.225 & 3.339 & 3.647 & 3.895(5)  \T\B\\

& 28 &1$^+$& 0.051  & 0.052    & 0.058 & 0.049 & 0.049 & 0.389(11)
& 2.229 & 2.124  & 2.253 & 2.208 & 2.080 & 2.426(3)  \T\B\\

& 29 &3/2$^+$& 0.115  & 0.107    & 0.079 & 0.102  & 0.079 & 0.085(3)
& 2.411 & 2.426  & 2.255 & 2.430 & 2.437 & 2.449(8)  \T\B\\

& 30 &2$^+$& -0.117  & -0.119 & -0.121 & -0.113  & -0.115 & 0.15(4) \cite{NNDC}
& 2.382 & 2.347  & 2.211 & 2.433 & 2.418 & 2.069(2) \cite{NNDC}  \T\B\\

& 31 &3/2$^+$& 0.076  & 0.069  & 0.044 & 0.068  & 0.058 & NA
& 2.549 & 2.604  & 2.568 & 2.595 & 2.614 & 2.305(8)  \T\B\\
\hline

\end{tabular}
\label{Moments}
\end{table*}
\end{landscape}
also the phenomenological USDB interaction reproduce the $B(E2)$ values within the experimental range for only selected transitions. The results are poor for medium mass Na isotopes ($^{25,26}$Na).

Table \ref{Moments} shows the quadrupole and magnetic moments for the g.s. of $^{20-31}$Na chain and also for the first excited states of $^{21,22,24}$Na. The experimental values are taken from Refs. \cite{NNDC,IAEA}.
The IMSRG interaction exactly reproduces the experimental quadrupole moment ($Q$) for the g.s. (2$^+$) of $^{20}$Na. The quadrupole moments of this state obtained from DJ16A and USDB interactions also agree reasonably with the experimental data. At the same time, the N3LO and JISP16 interactions predict slightly higher values than experimental data. For g.s. (3/2$^+$) of $^{21}$Na, the quadrupole moments obtained from all effective interactions, including USDB, are close to each other and smaller than the experimental value. The quadrupole moments calculated for the g.s. 3$^+$ of $^{22}$Na exhibit the same pattern, but their values are larger than the experimental data. All the interactions predict an oblate shape for the first excited states in $^{21}$Na and $^{22}$Na, for which the experimental data are not available. We obtained an enhanced value of quadrupole moment for the g.s. of $^{23}$Na from all interactions. We have calculated the quadrupole moments of the ground and the first excited state of $^{24}$Na, which do not have experimental data. The experimentalists may find this data useful for further studies on $^{24}$Na. The N3LO and JISP16 interactions failed to reproduce the correct sign of the g.s. quadrupole moment in $^{25}$Na. While only USDB interaction correctly reproduces the g.s. quadrupole moment in $^{26}$Na, the g.s. quadrupole moment of $^{27}$Na is exactly reproduced by only IMSRG interaction. A very small value of quadrupole moment is obtained from all interactions as compared to the experimental value for the g.s. of $^{28}$Na. Both USDB and DJ16A predict the same and slightly smaller $Q$ value than the experimental data for the g.s. of $^{29}$Na, while we obtained a larger $Q$ value from other interactions. All interactions failed to reproduce the observed sign of quadrupole moment in $^{30}$Na. The quadrupole moment of the g.s. of $^{31}$Na is not known yet experimentally. All the microscopic and USDB effective interactions predict a prolate configuration for this state.

The microscopic effective interactions overestimate the $Q$ values in $^{22}$Na. We observe as the number of neutrons increases (from $A=$ 25 to 27), the nuclei approach to nearly spherical shape. The neutrons occupy closed sub-shell configurations in $^{25,27,31}$Na. So both the experiment and the effective interactions predict weak collective properties in these nuclei. The effective single particle quadrupole moment of $^{25,27}$Na, corresponding to the last unpaired proton in d$_{5/2}$ orbitals, are 0.063 and 0.066 eb, respectively, which are considerably different from both the findings of the experiment and the shell model results. In Ref. \cite{em_properties_sd_shell}, the quadrupole moments of Na isotopes have been calculated using a mass-dependent VS-IMSRG interaction \cite{imsrg_A_dependent}. In our work, we have used an improved version of VS-IMSRG interaction \cite{imsrg_Z_dependent}, which is nucleus dependent and developed by the same group, but we found that the results have not improved significantly. All interactions produce poor results for neutron-rich isotopes $^{28-30}$Na. These nuclei lie inside or at the borders of the island of inversion region. So the quadrupole moments of these nuclei are affected by orbital configurations from higher model space and may be well described by microscopic effective interactions spanning $sdpf$- model space.

The g.s. magnetic moments of $^{20-22}$Na are overestimated by all interactions, but the overestimation is more in the case of N3LO and JISP16.  The N3LO and JISP16 gave better results for the first excited state of $^{21}$Na than other interactions, while the first excited state of $^{22}$Na is well described by DJ16A interaction as compared to other interactions. We obtained $\mu=2.209$ $\mu_N$ with DJ16A interaction for $^{23}$Na, which is close to the experimental value. In $^{24}$Na, all the microscopic interactions predict smaller $\mu$ values for the g.s. than the experimental data, while only USDB and IMSRG interaction reproduce the observed sign of magnetic moment of the first excited state. We observed that all microscopic effective interactions poorly reproduce the magnetic moments in $^{25-28}$Na. The calculated single particle magnetic moments for $^{25,27}$Na also differ significantly from the experimental data and shell model results. 
Except for DJ16A, the g.s. magnetic moment obtained for $^{29}$Na, from all interactions, are in reasonable agreement with the experimental data.
All interactions show larger values for magnetic moments in $^{30,31}$Na as compared to experimental data.   
\subsection{{\bf Rms radii and neutron skin thickness of Na isotopes}}\label{subsection3.4}
This section contains a discussion on the charge and matter radii as well as the neutron skin thickness calculated for Na isotopes from $A=$ 20 to $A=$ 31.

\subsubsection{{\bf Charge radii}}
The rms charge radius (R$_{ch}$) is one of the fundamental properties of the nucleus and can be measured accurately from experiments. The mean square charge radius ($\langle r_{ch}^2 \rangle $) of a nucleus is related to the mean square point proton radius ($\langle r_{pp}^2 \rangle $) by the expression given in Eq. \eqref{eq:6} \cite{global_formula}
\begin{equation}\label{eq:6}
\langle r_{ch}^2 \rangle = \langle r_{pp}^2 \rangle + \langle r_{p}^2 \rangle + \frac{N}{Z} \langle r_{n}^2 \rangle+ \frac{3\hbar^2}{4m_p^2c^2}.
\end{equation} 
Where R$_{ch}=\sqrt{\langle r_{ch}^2 \rangle}$, $\frac{3\hbar^2}{4m_p^2c^2}=0.033 fm^ 2$ (Darwin-Foldy correction term), $\langle r_{p}^2 \rangle$=  0.77$fm^2$ (mean square charge radius of proton) and $ \langle r_{n}^2 \rangle $= -0.1149$fm^2$ (mean square charge radius of neutron).

\begin{figure}
\begin{center}
\includegraphics[width=85mm]{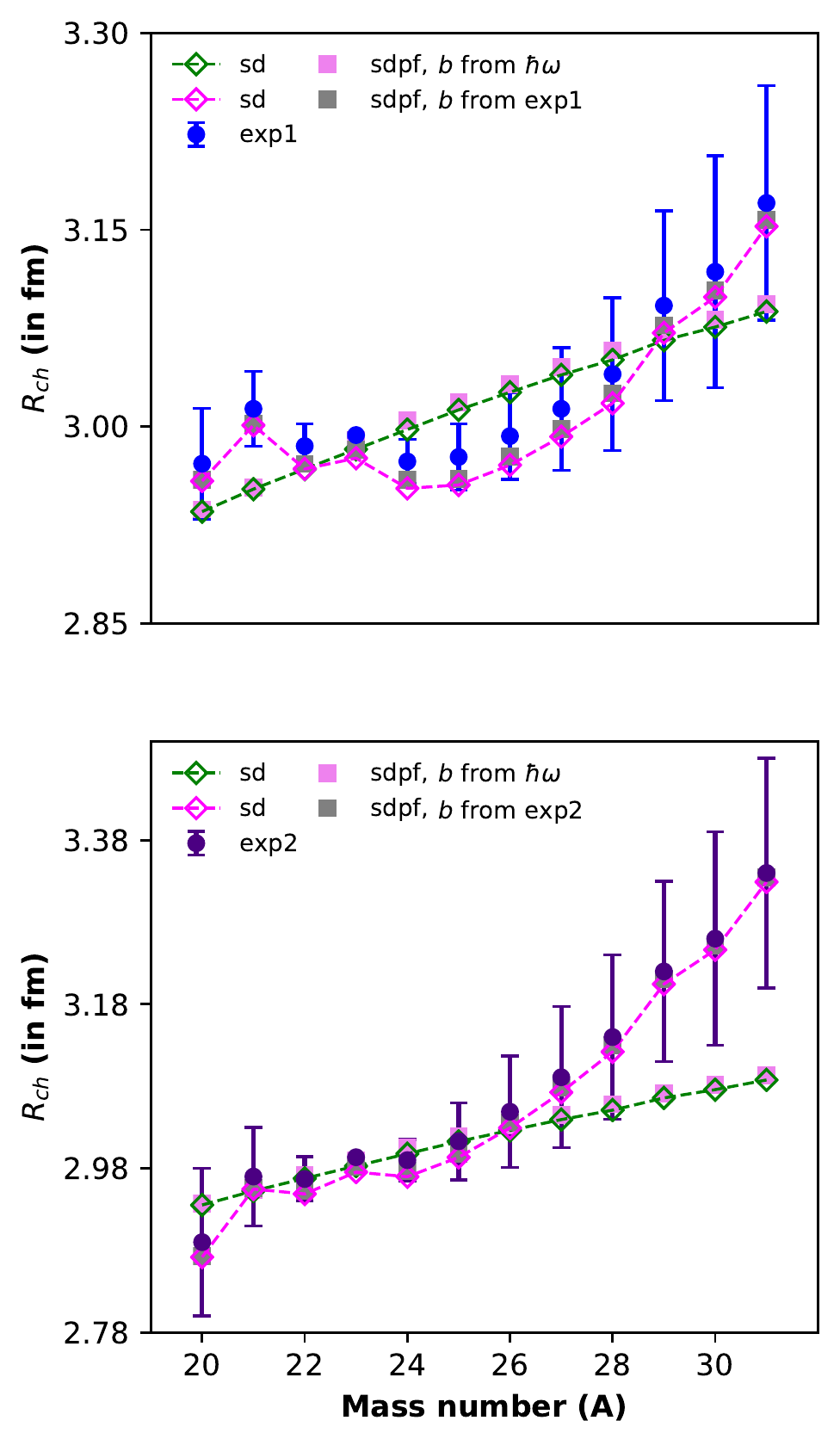}
\end{center}
\caption{ Charge radii of Na isotopes. Blue \cite{ch_radii_data} and indigo \cite{Na_atomic_and_nuclear_theory} symbols are for experimental data exp1 (top panel) and exp2 (bottom panel). The green (diamonds) and violet (squares) symbols refer to charge radii calculated for $sd$ and $sdpf$-shell respectively using $b$ from $\hbar\omega$. In other case $b$ is taken from the experimental data, for this 
magenta (diamonds) and gray (squares) symbols are charge radii obtained for $sd$ and $sdpf$-shell respectively.}
\label{fig:4}
\end{figure}

We have used the shell model H.O. basis states to calculate the point proton radii of Na isotopes. The point proton radii can be evaluated from proton occupation number of orbits and oscillator length parameter ($b$) \cite{Rp_N_isotopes}. Following this Ref. \cite{Rp_N_isotopes}, we have obtained the point proton radii of $^{20-31}$Na by considering the valence protons both in $sd$ and $sdpf$-shell {\color{black} as well as protons in the closed $^{16}$O core}. For oscillator length parameter ($b$), we have {\color{black}first} used $\hbar\omega= 45A^{-1/3}-25A^{-2/3}$. The shell model calculations are carried out with {\color{black} the microscopic effective interactions in $sd$-shell used in Sect. \ref{section2} and \ref{section3} as well as sdpf-mu interaction \cite{sdpfmu_interaction} to obtain point proton radii.} The point proton radii are then converted into charge radii using Eq. \eqref{eq:6}. {\color{black}Note that within the $sd$-shell, the charge radii do not depend on the effective interactions as the $s$- and $d$-orbits give the same radius in the H.O. basis and the point proton radii are determined by the sum of their occupation numbers}. In Fig. \ref{fig:4}, the calculated charge radii for $sd$- and $sdpf$-shell are compared with two sets of experimental charge radii data, one taken from Ref. \cite{ch_radii_data} (exp1) and another recently used in the literature \cite{Na_atomic_and_nuclear_theory} (exp2). From Fig. \ref{fig:4}, we can see that both the experimental data show a large increase in charge radii as $N$ exceeds 15. The charge radii calculated for $sd$-shell increase along the isotopic chain as $A$ increases. The addition of $fp$-shell {\color{black}up to 2$\hbar\omega$ excitations for the case of sdpf-mu interaction} has slightly enhanced the charge radii values. It is expected that the inclusion of more excitations (4 or 6$\hbar\omega$) in $fp$-shell will raise the charge radii values and will bring them closer to the experimental data in neutron excess Na isotopes. This implies that more configuration mixing in neutron-rich isotopes leads to an increase in nuclear charge radius as reported in \cite{Rm_Na_isotopes}. However, the experimental kink is not reproduced, and also there is a large deviation in calculated and experimental results for some isotopes. But the experimental charge radii are well reproduced when the parameter $b$ is extracted from the experimental charge radii data, as earlier reported in Ref. \cite{b_from_exp}. When $b$ is taken from the experimental data, the calculated charge radii are in excellent agreement with the observed data as shown in Fig. \ref{fig:4}.
This motivates us to use the same $b$ values for calculating matter radii and neutron skin in later sections.

\subsubsection{{\bf Matter radii}}
\begin{figure}
\begin{center}
	\includegraphics[width=85mm,height=70mm]{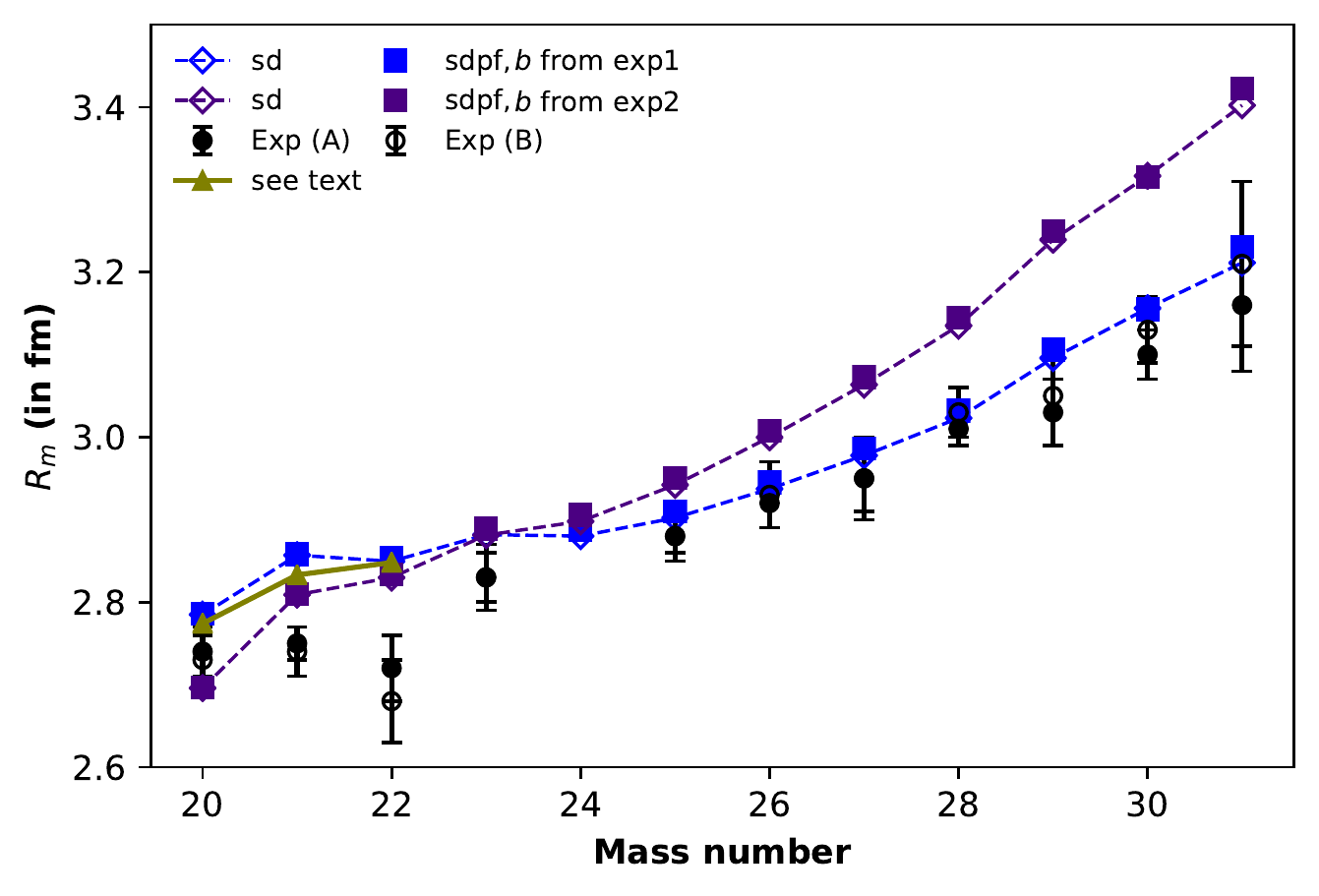} %
	\end{center}
		\caption{ Matter radii of Na isotopes. Filled (unfilled) black circles are for experimental data case A (B) \cite{Rm_Na_isotopes}. The blue and indigo symbols refer to matter radii calculated using $b$ from exp1 and exp2, respectively. Here diamond corresponds to calculated results for $sd$- shell and square for $sdpf$-shell.}
	\label{fig:5}
\end{figure}

The mean square matter radius is defined \cite{even_N_Na_moments} by Eq. \eqref{eq:7}
\begin{equation}\label{eq:7}
\langle r_{m}^2 \rangle =( Z\langle r_{pp}^2 \rangle + N\langle r_{pn}^2 \rangle)/A.
\end{equation}
Where $\langle r_{pn}^2 \rangle$ is the mean square point neutron radius. The calculation of point neutron radii are carried out in the same way as done for point proton radii for {\color{black} $sd$ and $sdpf$-shell as well as for $^{16}$O core.} Then the rms matter radii (R$_m$=$\sqrt{\langle r_{m}^2 \rangle}$) are obtained using Eq. \eqref{eq:7} for Na chain. In the literature \cite{Rm_Na_isotopes}, two sets of experimental matter radii values (A and B) have been reported for Na isotopes (except $^{24}$Na). We have compared our calculated matter radii with both experimental data A and B in Fig. \ref{fig:5}.

The observed matter distribution of the Na chain shows a gradual increase in matter radii as $A$ increases starting from $^{23}$Na. The calculated matter radii also follow the same trend. The matter radii obtained with $b$ values from exp1 data are in good agreement with the observed matter radii, while those obtained with $b$ values from exp2 data show larger values. 
We are getting the kink at $^{22}$Na, as observed in experimental data, when we have calculated matter radii using $b$ value obtained from exp1 data. The description about olive line at $A$ = 20 to 22 is given later in the text.

\subsubsection{{\bf Neutron skin}}
Neutron skin is a vital parameter in understanding the structural behaviour of a nucleus. It is defined as R$_{np}=$R$_{n}$-R$_{p}$ where $R_n = \sqrt{\langle r_{pn}^2 \rangle}$ and $R_p = \sqrt{\langle r_{pp}^2 \rangle}$.
\begin{figure}
\begin{center}
	\includegraphics[width=85mm,height=70mm]{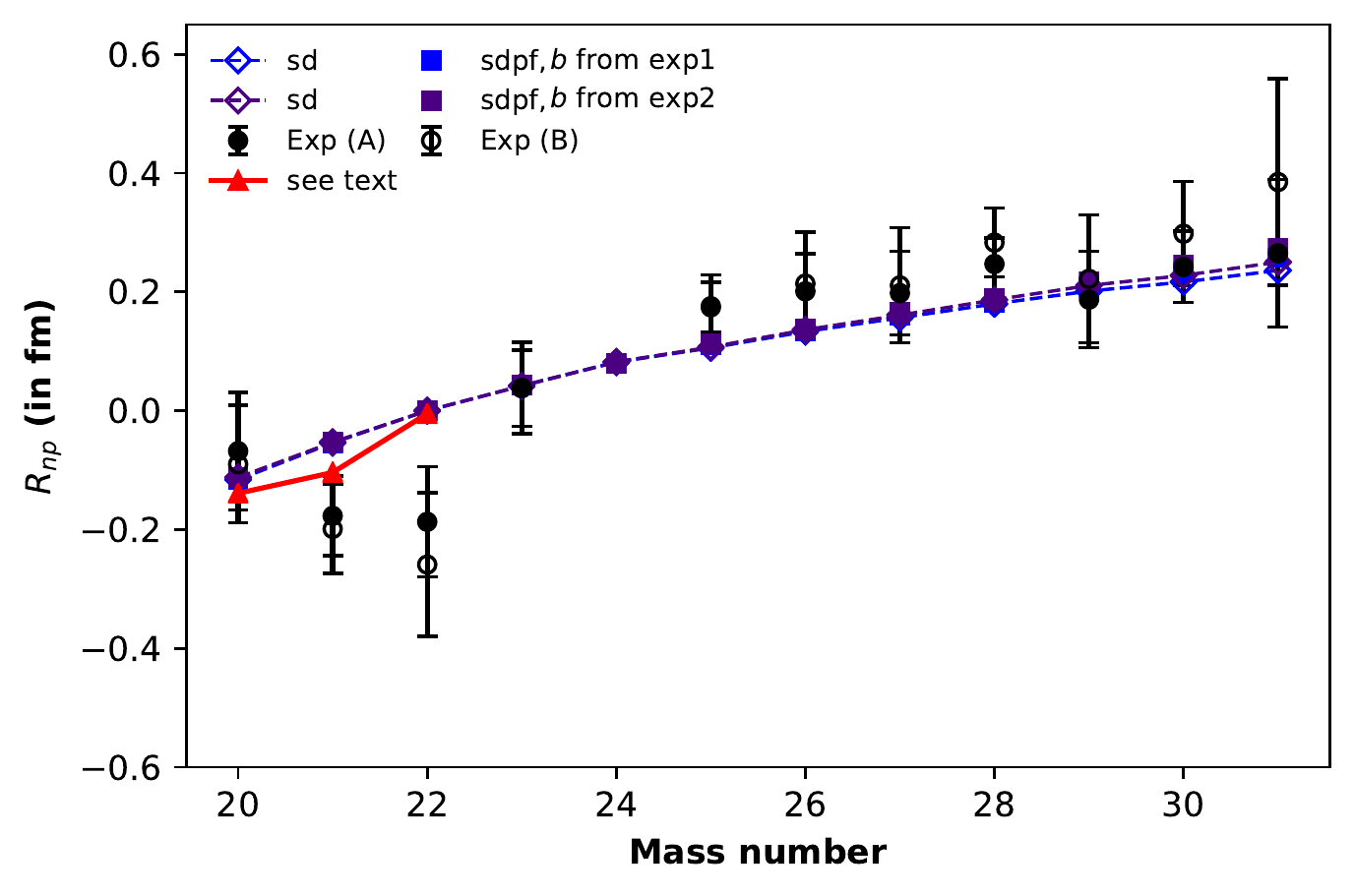} %
	\end{center}
	\caption{ Neutron skin thickness of Na isotopes. Filled (unfilled) black circles are for experimental data case A (B) \cite{exp_neutron_skin}. The blue and indigo symbols refer to neutron skins calculated using $b$ from exp1 and exp2, respectively. Here diamond corresponds to calculated results for $sd$- shell and square for $sdpf$-shell.}
	\label{fig:6}
\end{figure}
The calculated R$_{np}$ for $sd$ and $sdpf$-shell are shown in Fig. \ref{fig:6}. For experimental values, we have considered two possible sets of neutron skin for Na isotopes (similar to the case of matter radii) as reported in the literature \cite{exp_neutron_skin}. We find that the calculated neutron skins have almost the same value. In the observed neutron skins of the Na chain, $^{23}$Na has a very small positive neutron skin since it is stable and has one excess neutron. After that, the neutron skin gradually increases as $N$ increases along the chain. The calculated
neutron skins show the same trend and lie close to the experimental values. But in the neutron-deficient region ($^{20-22}$Na), the observed neutron skin behaves differently, i.e., with increase in number of neutrons, the neutron skin thickness is decreasing. Contray to this, our calculated R$_{np}$s show an increase in neutron skin thickness in this region.

We see that the calculated results for matter radii and neutron skin thickness are in good agreement with the experimental data for  $N \geq$ 12. However, the experimental behavior is not well reproduced in the neutron-deficient Na isotopes. The observed neutron skin is reduced in this region. This may be due to the effect of proton skin in these isotopes. The weak binding energies of protons in $^{20-22}$Na cause enhancement of proton radii while neutrons are deeply bound, and therefore, the neutron skin is more reduced at $A=$ 20 to 22.
This effect can be realized from the red curve in Fig. \ref{fig:6} at $A=$ 20 to 22. It connects the neutron skins calculated from R$_p$ (obtained from exp1 data) and R$_n$ (obtained using $b$ from $\hbar\omega$). A similar calculation has been done for matter radii at $A=$ 20 to 22 in Fig. \ref{fig:5} and represented by the olive line. We observed that the R$_m$ and R$_{np}$ values for $A=$ 20 and 21 come closer to the experimental data, but those at $A=$ 22 remain to be enhanced compared to the experimental value.
\section{Summary}
In the present work, we have done systematic shell model study of Na isotopes (from $A=$ 20 to $A=$ 31) in the $sd$-shell using different microscopic effective interactions. For few number of valence particles, all effective interactions are good in reproducing the observed g.s. energies, relative energy spectra and their spin-parities. Better results are obtained with monoploe modified interaction DJ16A. However, the low energy spectra of medium mass Na isotopes are not so well reproduced as in the case of lower mass Na nuclei. The neutron-rich Na isotopes ($^{29-31}$Na) lie in the island of inversion region and the shell model configurations are affected by intruder orbital configurations from $fp$-shell. The microscopic effective interactions overestimate the g.s. energy for such isotopes (except DJ16A which slightly underestimate the g.s. energy). 
Although, the observed spin-parities of g.s. and a few excited states in $^{29,31}$Na are correctly reproduced. The other nuclear properties, we studied, include reduced $E2$-transition strengths, quadrupole and magnetic moments. The microscopic effective interactions fairly reproduce the g.s. quadrupole and magnetic moments of the Na chain and also transition strengths for selected transitions. We have predicted these electromagnetic properties for some Na isotopes for which the experimental data is not available. In addition to it, the rms charge and matter radii, and neutron skin thickness of the Na chain are also discussed. For rms radii, the shell model calculations are carried out using H.O. wave functions in $sd$ and $sdpf$-shell. We found overall good agreement of rms charge, matter radii, and neutron skin with the observed data when $b$ is extracted from experimental charge data instead of obtaining it from $\hbar\omega$. The observed behavior of neutron skin thickness in the neutron-deficient region is better understood by taking account of the proton skin effect due to the weak binding energies of protons in these isotopes.
An explanation of the behavior at $A=$ 22 still remains a challenge.

\section{Acknowledgment}
S.S. would like to thank UGC (University Grant Com-
mission), India for financial support for his Ph.D. thesis
work. P.C.S. acknowledges a research grant from SERB
(India), CRG/2019/000556.

\end{document}